\documentclass[aps,pra,twocolumn,groupedaddress]{revtex4}

\usepackage{graphicx}
\usepackage{bm}

\begin{document}


\title{Using Absorption Imaging to Study Ion Dynamics in an Ultracold Neutral Plasma }


\author{ C. E. Simien, Y. C. Chen, P. Gupta,  S. Laha,  Y. N. Martinez,
P. G. Mickelson, S. B. Nagel, and T. C. Killian}
\affiliation{Rice University, Department of Physics and Astronomy
and Rice Quantum Institute, Houston, Texas, 77251}


\date{\today}

\begin{abstract}
We report optical absorption imaging of ultracold neutral plasmas.
Images are used to measure
the ion absorption spectrum, which is Doppler-broadened. Through
the spectral width, we monitor ion equilibration in the first 250
ns after plasma formation. The equilibration leaves ions on the
border between the weakly coupled gaseous and strongly coupled
liquid states. On a longer timescale of microseconds, we observe
radial acceleration of ions resulting from pressure exerted by the
trapped electron gas.
\end{abstract}


\maketitle
 Plasma physics traditionally studies systems with temperatures of thousands of
 kelvin or greater because  collisional ionization of atoms requires
 kinetic energies on this scale. Ultracold neutral plasmas,
 created by photoionizing
 laser-cooled and trapped atoms, access an exotic regime in which
 particle energies can be on the order of 1 K.

 Fundamental interest in ultracold neutral plasmas stems from a
range of phenomena in the ultracold regime.
Recent experiments have studied plasma creation \cite{kkb99},
collective modes \cite{kkb00}, and recombination into highly
excited Rydberg atomic states \cite{klk01}. The spontaneous
evolution of a dense, ultracold gas of Rydberg atoms into an
ultracold plasma has also been investigated \cite{rtn00}. This is
part of a search in atomic clouds for an analog of the Mott
insulator-conductor transition in condensed matter \cite{vrg82}.
Recombination in these systems resembles the methods used to
produce cold antihydrogen with trapped positrons and antiprotons
\cite{aab02, gbo02}.

A series of theory papers \cite{mur01,kon02,mck02,rha03,tya00}
explored issues surrounding thermalization and recombination in
ultracold neutral plasmas when both electrons and ions are near or
in the strongly coupled regime \cite{ich82}. In strongly coupled
plasmas the electrical interaction energy between the charged
particles exceeds the average kinetic energy. This reverses the
traditional energy hierarchy that underlies our normal
understanding of plasmas based on concepts such as Debye screening
and hydrodynamics. Strongly coupled plasmas exist in dense
astrophysical systems \cite{vho91},  matter irradiated with
intense laser fields \cite{nmg98, sht00}, dusty plasmas of highly
charged macroscopic particles \cite{mtk99}, and non-neutral
trapped ion plasmas \cite{mbh99} that are
 laser cooled until they freeze into Wigner crystals.

We now report the first results with a new probe of ultracold
plasmas: absorption imaging. This technique 
provides {\it in situ}, non destructive measurements, and offers
excellent spatial, temporal, and spectral resolution. We describe
the use of this probe to study ion-ion equilibration and expansion
of the plasma during the first few microseconds after
photoionization, but we emphasize its great potential to study a
host of phenomena such
as ion collective modes \cite{mur00}, 
shock waves \cite{rha03}, recombination, and particle-particle
spatial correlations \cite{mbh99}.

\begin{figure}
  \includegraphics[width=3in,clip=true,trim=50 250 10 80 ]{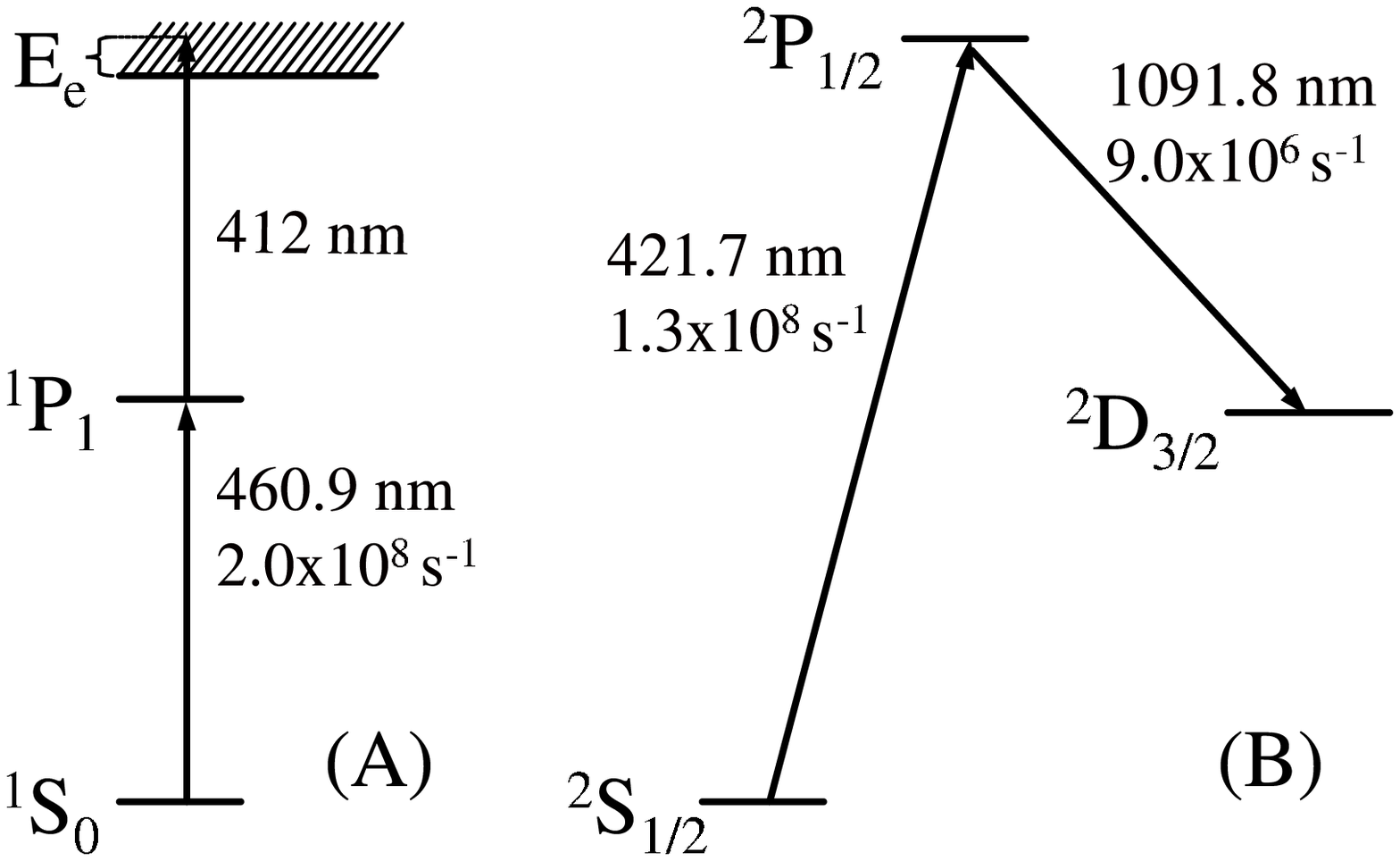}
  \caption{Atomic and ionic energy levels involved in the experiment, with decay rates.
  (A) Neutral atoms are laser cooled and trapped in a magneto-optical trap (MOT) operating on the
   $^1S_0-{^1P_1}$ transition at 461 nm.
Atoms  excited to the $^1P_1$ level by the MOT lasers are ionized
by photons from a laser at $\sim 412$~nm.
 (B) Ions are imaged using the $^2S_{1/2}-{^2P_{1/2}}$ transition at $422$~nm.
 $^2P_{1/2}$ ions decay to the $^2D_{3/2}$ state 7\% of the time, after which
 they cease to interact with the imaging beam.
This does not complicate the experiment because ions typically
scatter
 fewer than one photon during the time the imaging beam is on.}
 \label{energy}
\end{figure}

The production of an ultracold neutral plasma starts with atoms
that are cooled and confined in a magneto-optical trap (MOT)
(Fig.\ 1A). This aspect of the experiment was described in
\cite{nsl03}. The neutral atom cloud is characterized by a
temperature of about 20 mK and a density distribution given by
$n({r})=n_0{\rm exp}(-r^2/2\sigma^2)$, with $\sigma = 1$~mm and
$n_0=(4 \pm 2) \times 10^{10}$~cm$^{-3}$. The number of trapped
atoms is $(6 \pm 1) \times 10^8$.

To form the plasma, the MOT magnets are turned off and atoms are
ionized with photons from a $10$~ns pulsed dye laser whose
wavelength is tuned just above the ionization continuum (Fig.\
1A). Because of the small electron-to-ion mass ratio, the initial
electron kinetic energy ($E_e$) approximately equals the
difference between the photon energy and the ionization potential.
$E_e/k_B$ can be as low as the bandwidth of the ionizing laser,
which is $\sim$100 mK. The initial kinetic energy for the
resulting singly-charged, electronic ground state ions is close to
that of the original neutral atoms. As we will discuss below, the
resulting non-equilibrium plasma evolves rapidly. Up to $12 \pm
1$\% of the neutral atoms are ionized producing plasmas with a
peak density of $(5 \pm 3) \times 10^{9}$~cm$^{-3}$.

Immediately after photoionization, the charge distribution is
neutral everywhere. Due to the kinetic energy of the electrons,
the electron cloud expands on the timescale of the inverse
electron plasma frequency $\tau_e=\omega_{pe}^{-1}=\sqrt{m_e
\varepsilon_0/ n_{e} e^2}< 1$~ns, where $m_e$, $n_e$, and $e$ are
the electron mass, density and charge. On this timescale the ions
are essentially immobile. The resulting local charge imbalance
creates a Coulomb potential energy well that traps all but a small
fraction ($<5$\%) of the electrons. Simulations \cite{kkb99} show
that electrons escape mostly from the edges of the spatial
distribution, and the center of the cloud is well described as a
neutral plasma \cite{nonneutralnote}. The diagnostic used in
previous experiments was charged particle detection of electrons
and ions after they had left the plasma.


Spectroscopic diagnostics are ubiquitous in plasma experiments,
and some even provide spatial information, such as spatially
resolved laser induced fluorescence \cite{ltr01, mbh99}. The
absorption imaging reported here is particularly well-adapted for
small, cold, relatively dilute plasmas that evolve very quickly.
It is also a powerful technique for studying laser cooled and
trapped neutral atoms \cite{mvs99}. A collimated laser beam, tuned
near resonance with the principle transition in the ions (Fig.\
1B), illuminates the plasma and falls on an image intensified CCD
camera. Following Beer's law, the optical depth ($OD$) is defined
in terms of the image intensity without ($I_{background}$) and
with ($I_{plasma}$) the plasma present,
\begin{eqnarray}\label{OD}
OD(x,y)&=&{\rm ln}(I_{background}(x,y)/I_{plasma}(x,y)) \nonumber \\
       &=&\alpha(\nu)\int_{-\infty}^{\infty} dz \hspace{.025in}
       n_i(x,y,z), \nonumber \\
       &=&{{n_{0i} \alpha(\nu)
}\over{\sqrt{2 \pi} \sigma_z}}{\rm
e}^{-x^2/2\sigma_x^2-y^2/2\sigma_y^2}
\end{eqnarray}
 where
 $n_{0i}$ is the peak ion density, and $\alpha(\nu)$ is the absorption cross section at
 the image beam
frequency, $\nu$. In the last line we have inserted a Gaussian
density distribution for the ions, which leads to the function
used to fit the data.


\begin{figure}
  \includegraphics[width=3.3in,clip=true]{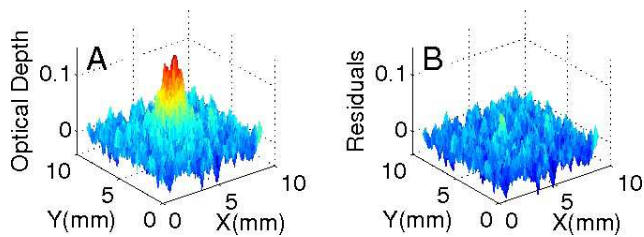}\\
  \caption{Optical depth of an ultracold neutral plasma.
  (A) The delay between the formation of the plasma and
image exposure is $100$~ns, and the initial
  peak density is $n_{0i}=5 \times 10^{9}$~cm$^{-3}$.
(B) Residuals of a fit to a two-dimensional Gaussian profile are
close to the photon shot-noise. }\label{image}
\end{figure}

Figure \ref{image} shows a typical absorption image. The intensity
of the probe beam is approximately $200$~$\mu {\rm W}/{\rm cm}^2$,
which is much less than the saturation intensity of the transition
($38$~$ {\rm mW}/{\rm cm}^2$).
The spatial resolution of typical images is about
$100$~$\mu$m, limited by pixel averaging performed to improve the
signal-to-noise ratio of the images.

To study the time-evolution of the plasma, we vary the delay
between the formation of the plasma and image exposure
($t_{delay}$) with $10$~ns accuracy. The minimum camera exposure
gate width is  $50$ ns.
For the shortest exposure times, which we use for the best time
resolution at very short $t_{delay}$, we typically average about
60 ionizing laser shots. For longer delay times we use longer
image exposure times of up to $800$~ns, and decrease the number of
accumulations to keep the total number of photons detected
approximately constant. The repetition rate for ionization and
image recording is about 5 times per second.

The imaging beam itself can also be gated with a minimum width of
about $200$~ns. We use this capability for $t_{delay}> 1$~$\mu$s
to turn off the image beam until the camera exposure begins. This
prevents optical pumping of the ions to the $^2D_{3/2}$ state
(Fig. 1B).

\begin{figure}[b]
  \includegraphics[width=3in,clip=true]{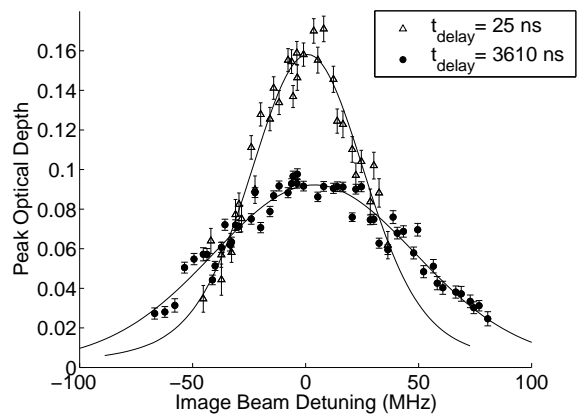}\\
  \caption{Absorption spectra of ultracold neutral plasmas.
  We plot the peak optical depth derived from fits to data such as Fig.\ 2.
  The frequency is with respect to a Doppler-free absorption
  feature in a strontium discharge cell. Both spectra correspond to $E_e/k_B=68$~K and
  the same initial peak plasma density of $n_{0i}=5 \times
  10^{9}$~cm$^{-3}$.  Data are fit with Voigt profiles, and the increase in
  linewidth for longer $t_{delay}$ is clear.}\label{spectrum}
\end{figure}

Plotting the peak optical depth as a function of image laser
frequency provides the absorption spectrum of the ions (Fig.\
\ref{spectrum}). The imaging laser linewidth of about $5$~MHz  is
negligible on the scale of the $21.5$~MHz natural linewidth of the
transition. As described below, additional broadening of the
absorption spectrum provides a wealth of information on the plasma
dynamics.

Here we describe experiments using the time evolution of the
absorption spectrum to study ion dynamics
for a plasma with $N_i=7 \times 10^7$ ions, initial peak density
for ions and electrons of $n_{0i}\approx n_{0e}=(5 \pm 3) \times
10^9$~cm$^{-3}$, and $E_e/k_B=(68 \pm 5)$~K. We chose a relatively
large $E_e$ in order to avoid complications that arise when the
electron Coulomb coupling parameter ($ \Gamma_e = e^{2} / 4\pi
\varepsilon_{0}\,a k_{B}T_e$) approaches or initially exceeds
unity, such as screening of the ion interaction \cite{fha94}, and
rapid collisional recombination and heating of the electrons
\cite{kon02,mck02,rha03,tya00}. Here, $a=(4\pi n_{0e}/3)^{-1/3}$
is the Wigner-Seitz radius, and $T_e={2 \over 3} E_e/k_B$ is the
electron temperature as set by the wavelength of the ionizing
laser. For this sample $ \Gamma_e =0.1 $.

\begin{figure}
  \includegraphics[width=3.0in,clip=true]{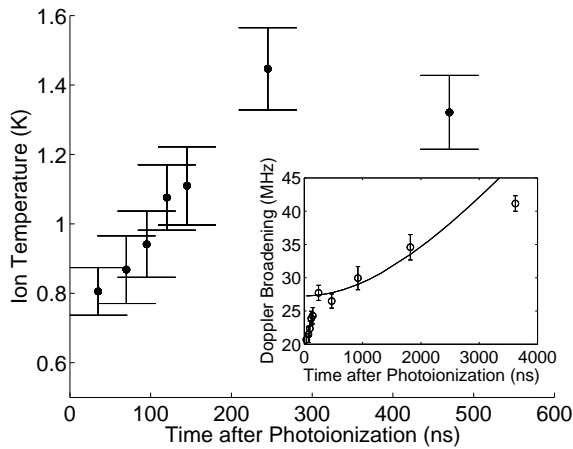}\\
  \caption{Ion dynamics determined from the spectral width.
  Error bars are two sigma uncertainties from Voigt profile
  fits to absorption spectra. The time after photoionization
  corresponds to the
  timing of the center of the camera exposure
  gate. Data show disorder-induced
  heating of the ions as the ions equilibrate during
  $t_{delay}<250$~ns. (Inset) For longer times,
  the Doppler broadening of the spectrum continues to increase as
  ions accelerate radially due to electron
  pressure. This process is described by a hydrodynamic model with no
  free parameters.
  }\label{tempevolution}
\end{figure}

The observed spectral linewidths are all significantly broader
than the natural linewidth of the transition.  Collisional
broadening is not significant because of the relatively low
particle density and collision frequency in the sample. The
dominant contribution to the linewidth is Doppler broadening,
which makes the spectrum a very accurate probe of the ion velocity
distribution. From the fit of each spectrum to a Voigt profile, we
extract the rms Doppler broadening, $\sigma_D={\sqrt{k_B
T_i/m_i}\over \lambda}$, where  $m_i$ is the ion mass, and
$\lambda$ is the wavelength of the transition. This allows us to
determine an ion temperature ($T_i$) as a function of time, as
shown in Fig.\ \ref{tempevolution}.

The temperature increases rapidly for $t_{delay}<250$~ns. Two
pieces of information imply that this is thermalization of ions
with themselves after creation in a spatially disordered state.
The timescale is on the order of the inverse plasma frequency of
ions, $\tau_i=\omega_{pi}^{-1}=\sqrt{m_i \varepsilon_0/n_{0i}
e^2}= 100$~ns, which is the timescale on which ions respond to
perturbations from their equilibrium spatial distribution. The
temperature of $1.4 \pm 0.1$~K derived from the Doppler width at
$250$~ns is also on the order of the amount of potential energy
inherent in the initial disorder of the ions. A simple estimate
implies there is $e^2/4\pi \varepsilon_0 a k_B\approx5$~K of
potential energy that will be redistributed during thermalization.

 This thermalization has been modelled with molecular dynamics
 simulations \cite{mur01, mck02, kon02}, and the experimental data is in general
 agreement with the theory. The accuracy of the imaging probe
 will enable detailed comparison of measurement and theory for the time dependence of the
ion temperature.

The final temperature reached, however, can be compared with an
expression derived in \cite{mur01}.
 Assuming complete initial disorder and incorporating the screening effects of the
 electrons,
\begin{eqnarray}\label{iontemp}
  T_i&=&{2 \over 3} {e^2 \over
 4\pi \varepsilon_0 a k_B}\mid \tilde{U} +{\kappa \over
 2}\mid .
\end{eqnarray}
 Here, $\kappa =a/\lambda_D=\sqrt{3\Gamma_e}=0.55$ where
 $\lambda_D=(\varepsilon_0 k_B T_e/ n_{0e} e^2)^{1/2}=7$~$\mu$m is the Debye length. The
 quantity $\tilde{U}\equiv {U \over N_i e^2 /
 4\pi \varepsilon_0 a} $ is the potential energy per particle
 in units of $e^2/4\pi \varepsilon_0 a$. It has
 been studied with molecular dynamics simulations \cite{fha94}
 for a homogeneous system of particles interacting through a
 Yukawa
 potential, $\phi(r)= {e^2 \over 4\pi \varepsilon_0 r}{\rm
 exp}(-r/\lambda_D)$, which describes ions in the background of weakly
 coupled electrons \cite{debyespherenote}.
 The experimental values for $n_i$, $T_i$, and $\kappa$,
 imply  $\tilde{U} $=-0.73. Equation \ref{iontemp} then gives
$T_i=1.4$~K, in
 excellent agreement with the measured value.

We now address the level of Coulomb coupling for the ions. From
the temperature and peak density we derive $\Gamma_i=3 \pm 1$ for
the thermalized ion cloud at $t_{delay}=250$~ns. For a system of
charges embedded in a uniform neutralizing background, formally
called a one-component plasma \cite{don99}, local spatial
correlations characteristic of a strongly coupled fluid appear for
$\Gamma \ge 2$. For Debye shielded particles, such as the ions
studied here, screening will reduce the correlations. This effect
is approximately incorporated by using the effective coupling
constant $\Gamma^* = \Gamma{\rm e}^{-\kappa}$ \cite{kon02, fha94}.
At $250$~ns, $\Gamma_i^*=2 \pm 1$, and the ions are just on the
edge of the strongly coupled fluid phase. Perhaps other initial
experimental parameters, or laser cooling of the ions \cite{kag03,
kon02}, will lead to more strongly coupled systems, although
$\Gamma_i^*=2$ already puts the experiment in an interesting
regime.

For $t_{delay}>250$~ns, the spectral width continues to increase,
but at a slower rate, as shown in the inset of Fig.
\ref{tempevolution}. This slow increase is due to outward radial
acceleration of the ions caused by pressure exerted by the gas of
trapped electrons. This was studied experimentally in \cite{kkb00}
and theoretically by a variety of means in \cite{rha03}. The
experiments measured the final velocity that the ions acquired,
which was approximately $\sqrt{E_e/m_i}$. Here we observe the ion
dynamics at much earlier times during the acceleration phase.

A hydrodynamic treatment \cite{kkb00} predicts that the force per
ion is
\begin{eqnarray}\label{ionforce}
\bar{F}&=& {{\bar \nabla}(n_e(r) k_B T_e) \over n_i(r)} = \hat{r}
{r k_B T_e \over \sigma_i^2},
\end{eqnarray}
 where the ion and electron density distributions are
 $n_{e}(r)\approx n_{i}(r)=n_{0i}{\rm
 exp}(-r^2/2\sigma_i^2)$. We approximate the rms size ($\sigma_i$)
 as the mean of the observed sizes $\sigma_x$ and $\sigma_y$,
 and we assume thermal equilibrium for the electrons throughout the
 cloud \cite{rha03}. This force leads to a radial expansion velocity for
 the ions, $v_r(r)$,
 which is correlated with position and increases linearly with time.
 This does not represent an increase in the random thermal
 velocity spread or temperature of the ions.
 Due to the large mass difference, thermalization of ions and
 electrons \cite{kon02} is slow and occurs on a millisecond timescale.
 The increase in Doppler broadening due to thermalization is approximately one
 order of magnitude
 smaller than what is observed.

Using the dynamics implied by Eq.\
 \ref{ionforce}, the evolution of the Doppler broadening
can be calculated. The mean squared velocity component along the
imaging laser is
\begin{eqnarray}\label{ionvz}
&&\langle v_z^2\rangle= \int d^3r\, dv_T {n_i(r)\over N_i} P(v_T)
(v_T+v_r(r)cos\theta )^2,
\end{eqnarray}
 where $P(v_T)$ is the thermal distribution of $v_z$
 for $T_i=1.4$~K. The resulting theoretical Doppler broadening
$\sqrt{\langle v_z^2 \rangle}/\lambda$,
 plotted in Fig.\ \ref{tempevolution}, reproduces the data accurately.

The ion acceleration is thus an excellent diagnostic of the
electron temperature. This will be of great value in future
studies because the temperature is predicted to evolve in a
complicated fashion for lower initial $\Gamma_e$
 due to recombination and  disorder-induced heating of the
 electrons \cite{kon02,mck02,rha03,tya00}.
One also expects that there will be a cooling effect at longer
$t_{delay}$ due to expansion of the plasma and evaporative cooling
\cite{rha03}. The small discrepancy between theory and data for
long $t_{delay}$ in Fig.\ \ref{tempevolution} may indicate the
onset of this cooling, although expansion of the plasma is small
on the timescale of these observations. For the maximum
$t_{delay}$ and typical $v_r \approx 15$~m/s, $v_r
t_{delay}\approx 50$~$\mu$m. This small increase in size is
observed in the images (Fig.\ \ref{image}).

The initial study using absorption imaging of an ultracold neutral
plasma has probed ion dynamics in the first few microseconds after
photoionization. It revealed disorder-induced heating that was
predicted in \cite{mur01}, and showed that the ions equilibrate on
the boundary of the gas-liquid transition. Acceleration of ions
due to electron pressure was also evident, and can be used to
monitor the electron temperature.

Many future experiments suggest themselves. Some of the most
interesting are investigating dynamics when the initial electron
Coulomb coupling parameter is large and recombination and
disorder-induced electron heating  are expected to dominate the
plasma evolution. Detailed study of ion and electron
thermalization at the border of the strongly coupled regime is
also possible. Improvements in the imaging optics will
significantly increase the image signal-to-noise ratio and allow
study of features on the ion density distribution with $\sim
10$~$\mu$m experimental resolution.

This research was supported by the Department of Energy Office of
Fusion Energy Sciences, Office for Naval Research, Research
Corporation, Alfred P. Sloan Foundation, and David and Lucille
Packard Foundation.


\begin{thebibliography}{27}
\bibitem[{non()}]{nonneutralnote}
\bibinfo{note}{The approximate excess ion density is $3 n_e (\lambda_D /
  \sigma)^2 \ll n_e$, where $\lambda_D$ is the Debye screening length.}
\bibitem[{deb()}]{debyespherenote}
\bibinfo{note}{There are many electrons per Debye sphere ($\kappa^{-3}=n_e 4
  \pi \lambda_D^3/3=6$), which justifies the use of this model.}
\end{thebibliography}

\end{document}